# Stacking and Thickness Effects on Cross-Plane Thermal Conductivity of Hexagonal Boron Nitride

S. G. Wang[1+], H. F. Feng[1+], Zhi-Xin Guo[1, a)]

[1]State Key Laboratory for Mechanical Behavior of Materials, Center for Spintronics and Quantum System, School of Materials Science and Engineering, Xi'an Jiaotong University, Xi'an, Shaanxi, 710049, China.

Authors to whom correspondence should be addressed: zxguo08@xjtu.edu.cn;

**Abstract**

Recently, the in-plane thermal transport in van der Waals (vdW) materials such as graphene, hexagonal boron nitride (h-BN), and transition metal dichalcogenides (TMDs) has been widely studied. Whereas, the cross-plane one is far from sufficient. Based on the non-equilibrium molecular dynamics simulations and Boltzmann transport equation, here we reveal the stacking and thickness effects on the cross-plane thermal conductivity ($\kappa$) of h-BN. We find that $\kappa$ can be significantly modulated by both the stacking structure and thickness (d) of h-BN, which is unexpected from the viewpoint of its smooth in-plane structure and weak interlayer interaction. In the small thickness region (d<6 nm), $\kappa$ of h-BN at room temperature significantly increases with thickness, following a power law of $\kappa \propto d^\beta$ with $\beta$=0.84, 0.66, and 0.92 for AA', AB, and AB' stacking structures, respectively. Moreover, $\kappa$ of AB' structure reaches up to 60% larger than that of AA' and AB structures with d=5.3 nm, showing the remarkable stacking effect. We also find that the stacking effect on $\kappa$ changes dramatically with d increasing, where AA' stacking has the largest $\kappa$ with d > 200 nm. We finally clarify that such exotic stacking and thickness dependence of $\kappa$ is owing to the competing effect of excited number of phonons and phonon relaxation time, both of which directly affect the thermal conductivity. Our findings may provide new insights into the cross-plane thermal management in vdW materials.

# Introduction

In the past decade, thermal transport property in van der Waals (vdW) materials has attracted significant attention[1–8]. Until now, the in-plane thermal conductivity ($\kappa$) in many monolayer and multilayer vdW materials, such as graphene, black phosphorus, and TMDs, etc., has been well explored. Many exotic thermal transport phenomena have been discovered, including the unusually large phonon lifetime of ZA mode in graphene[9], the outstanding anisotropy of $\kappa$ in black phosphorus[10], as well the dramatic interlayer coupling effect on $\kappa$ of multilayer graphene[11,12] and TMDs[13,14]. Whereas, the study on cross-plane $\kappa$ of vdW materials is far from sufficient.

Generally, the in-plane $\kappa$ is expected to be hardly affected by the stacking structure due to the long-range vdW interlayer interaction, which forms a weak and smooth potential for the in-plane vibrations. Nevertheless, our recent studies show that vdW interlayer interaction can induce strong stacking-dependent cross-plane vibrations, which leads to a very remarkable influence on the cross-plane $\kappa$ [15]. Therefore, a systemic investigation on the stacking structure dependence of cross-plane $\kappa$ is in greatly desirable. On the other hand, it had been widely obtained in low-dimensional materials with strong chemically bonded interaction that the thermal conductivity dramatically increases with the length L, flowing a power law of $\kappa \propto L^\beta$, where the nonzero $\beta$ is a result of ballistic phonon transport[16]. It is also an interesting topic to explore if such power law exists in a 3D material with weak vdW bonded interaction, and how it changes with stacking structure.

In this work, based on the well-known vdW material, hexagonal boron nitride (h-BN), we explore the stacking and thickness effects on cross-plane thermal transport properties. Note that the high in-plane thermal conductivity, atomically smooth surface, and large electronic band gap make h-BN an ideal substrate for 2D electronics[18–21]. Hence the clarification of its cross-plane thermal transport property will be helpful to the design of high-performance 2D electronic devices.

# Method

We consider three stable high-symmetry h-BN stacking sequences, namely AA'

stacking, AB stacking, and AB' stacking, as shown in Figs. 1(a)-(c). The stability of all three stacking structures has been proved by first-principles calculations[17]. The three structures can be also transformed into each other by rotating or sliding the basal plane.

The thermal conductivity is calculated by using the NEMD method, which is realized in the LAMMPS software package[22]. The optimized Tersoff potential[23] is used to describe the intralayer interaction. As for the vdW interlayer interaction, the ILP potential[24] is adopted with a cutoff interaction radius of 10 Å.

In NEMD simulations, the periodic boundary condition is applied to the in-plane x and y directions, and the fixed boundary condition is applied to the cross-plane z direction, respectively. The simulated box in the x and y directions are 4.96 nm and 4.29 nm, respectively. As shown in Fig. 2(a), the outmost layers at each side of the simulated box in the z direction are fixed. Then two nearby layers of each side are put into contact with the Langevin thermostat, which are called heat source and heat sink regions, respectively. The region between the heat source region and the heat sink region is denoted as the effective heat transfer region.

In the NEMD simulations, a time step of 0.5 fs is adopted. After structure optimization, the system is heated to a target temperature T (e.g. 300 K) by 200 ps ($4 \times 10^5$ step) with Nosé−Hoover thermostat. Then the heat source region and sink region are coupled with the Langevin thermostat by temperature $T_h$ and $T_c$, respectively. Note that $T_h = T + T_0$ and $T_c = T - T_0$, where $T_0$ is fixed at 30 K, and T corresponds to the average temperature of the system in the non-equilibrium steady state. All results given in this paper are obtained by averaging about 5 ns after a sufficiently long time 5 ns to set up a non-equilibrium stationary state. As shown in Fig. 2(b), due to the conservation of total energy in the system, the energy exchange rates of the heat source and heat sink remain basically the same at the stationary state.

The cross-plane $\kappa$ of h-BN in different stacking structures is calculated based on Fourier's Law,

$$\kappa = \frac{J}{|\nabla T|} = \frac{J}{\Delta T/d}$$

where J denotes heat flux per unit area, $\Delta T = 2T_0$ is the temperature difference between the heat source and heat sink regions, and d is the length of the thickness of the h-BN film.

On the other hand, $\kappa$ of h-BN can be obtained by solving the phonon Boltzmann transport equation (BTE), as implemented in the ShengBTE package[25]. In order to make direct compassion with the NEMD simulations, which are based on the empirical potentials, we have developed a code to bridge LAMMPS and ShengBTE. Using the code, the second-order interatomic force constants can be completely derived based on LAMMPS, which is much faster than the DFT calculations. The third-order interatomic force constants are investigated by ShengBTE combined with the finite displacement method using a $3 \times 3 \times 3$ supercell. In the ShengBTE calculations, a relaxation time approximation (RTA) method is used to solve the Boltzmann equation. A $20 \times 20 \times 20$ Monkhorst-Pack q-point mesh is adopted for the convergence on the summation. In addition, Phonopy[26,27] is used to calculate phonon bands and phonon density of states, etc.

## Results

We first explore the thickness d dependence of cross-plane $\kappa$ with different stacking structures in the small thickness region (d<6 nm). The results are obtained by using the NEMD method, which is particularly suitable for the evaluation of finite size effect[28–30]. Fig. 3(a) shows the calculated $\kappa$ of h-BN for all three stacking structures at room temperature (300 K). A common feature is that $\kappa$ significantly increases with thickness, following a power law of $\kappa \propto d^\beta$ with $\beta$=0.84, 0.66, and 0.92 for AA', AB, and AB' stacking structures, respectively. Note that similar size-dependent $\kappa$ had been observed in low-dimensional systems with strong covalent interactions, where $\beta$ usually varies around 0.5 at room temperature[30–32]. The index β reflects the weight of ballistic phonon transport (BPT) in the system. Particularly, β = 1 means that only BPT exists and the phonon mean free path (PMF) is infinite, whereas β = 0 indicates the size of the thermal conductor is much larger than the PMF, and no BPT exists in the system. Therefore, the relatively large β values obtained in the small thickness region show the strong BPT characteristic in the cross-plane direction, which is from the very weak vdW interlayer interaction.

Another interesting phenomenon is the strong stacking structure dependence of $\kappa$. As shown in Fig. 3(a), $\kappa$ of AB' stacking is obviously larger than that of AA' and AB stacking, which becomes increasingly significant with thickness increasing. For instance, $\kappa$ of AB' structure reaches up to 60% larger than that of AA' and AB structures with d=5.3 nm. Note that $\kappa$ of AA' stacking structure with d=5.3 nm is about 0.26 $W \cdot m^{-1} \cdot K^{-1}$, which agrees well with the recent experimental observation[33]. This result confirms the validity of our NEMD simulations.

Nevertheless, the ILP potential adopted in this study leads to unusually heavy computational resources and thus makes the NEMD method inefficient for the calculation of $\kappa$ of h-BN with larger thickness due to the large number of atoms in a simulated box (800 atoms per layer). To get an entire evolution of thermal conductivity with thickness, we further calculate thermal conductivity by using the Boltzmann transport equation method, which is implemented in the ShengBTE software. In order to make direct compassion with the NEMD results, we have developed a code to bridge LAMMPS and ShengBTE, where the second-order interatomic force constants can be completely derived based on LAMMPS. Although ShengBTE cannot directly consider the finite thickness effect due to its requirement on the periodic boundary condition for the continuous wave vector, it can still give rise to reasonable results if the boundary scattering is not remarkable. In this case, the finite thickness effect can be evaluated by changing the cutoff PMF (equivalent to thickness d) during the calculation.

Fig. 3(b) shows the PMF dependence of $\kappa$ for all the stacking structures obtained from the BTE method at 300 K. As shown in the inset, the BTE calculation gives rise to similar thickness dependence of $\kappa$ to the NEMD simulation in the small thickness region (d<10 nm). That is, $\kappa$ of all h-BN structures are all smaller than 1.0 $W \cdot m^{-1} \cdot K^{-1}$, and the AB' stacking structure has obviously larger $\kappa$ than the AA' and AB structures. This result confirms the validity of evaluating the size effect of $\kappa$ by using the cutoff PMF in the BTE method[34]. It is noticed that, as the thickness further increases to about 30 nm, the slope of $\kappa$ with thickness in AB' structure becomes smaller than that of AA' and AB structures. As a result, a crossover of $\kappa$ in the three structures appears when the thickness reaches about 200 nm. Interestingly, the converged thermal conductivities can be obtained only when the thickness reaches up to 1000 nm, showing the unusually large PMF originated from the weak vdW interactions. It is noteworthy that the AA' structure has the largest $\kappa$ of 5.2 $W \cdot m^{-1} \cdot K^{-1}$, which is about 24% and 30% larger than that of AB (4.2 $W \cdot m^{-1} \cdot K^{-1}$) and AB' structures (4 $W \cdot m^{-1} \cdot K^{-1}$), respectively. This result is in contrast to the stacking structure dependence of $\kappa$ in the small thickness region.

To understand the underlying mechanism for the complex thickness-dependent sacking effect on $\kappa$ of h-BN, we further calculate the phonon band structures. As shown in Figs. 1(d)-1(f), the phonon branches in the in-plane directions (Γ-M-K-Γ paths) are almost the same for the three stacking structures, indicating the robustness of in-plane $\kappa$ against the stacking structure. This result can be well understood from the long-range vdW interlayer interaction, which provides a weak and smooth potential to the in-plane vibrations. As for the cross-plane direction (Γ-A path), however, there is obvious stacking-structure dependent low-frequency transverse phonon modes (TA and TO'). In comparison with the phonon bands of the AA' structure, the TA and TO' bands slightly upshift in the AB structure, whereas they significantly downshift in the AB' structure. As discussed below, the variation of TA and TO' bands can result in distinct phonon-phonon scatterings in the low frequency region ($\omega$<5 THz), and thus the strong stacking-structure dependent cross-plane $\kappa$.

Figs. 4(a)-4(c) additionally show the projected cross-plane $\kappa$ of h-BN on the basal plane (0 0 1) in the Brillouin zone (BZ) based on the ShengBTE calculations. It is seen that in all three h-BN structures, cross-plane $\kappa$ is mostly concentrated around the Γ point in the 2D BZ. This feature shows that the cross-plane thermal transport is mostly contributed by the phonon modes along the cross-plane Γ–A path. Hence, in the following discussions, we focus on the contributions of phonon modes along Γ–A path to the cross-plane $\kappa$. Fig. 3(d) shows the relative contributions of LA, TA, and all the optical phonon modes along the Γ–A path. It is seen that $\kappa$ of the three stacking structures exhibits different phonon-mode-contribution characteristics. In AB' structure,

the contribution (84.8%) of acoustic phonon modes (TA, LA) is obviously smaller than that of AA' (89.5%) and AB (92.9%) structures. This result implies that the downshift (upshift) of TO' in AB' (AB) structure in comparation with that of AA' structure makes the optical phonon modes participate more (less) in the thermal conduction.

The calculated frequency dependence of $\kappa$ in Fig. 5(a) further shows that the cross-plane thermal conductivity is mainly contributed by the low frequency phonons ($\omega$<5 THz). This result confirms that the upshift/downshift of TA and TO' phonon bands that are in the low energy region can lead to significant variation of cross-plane $\kappa$. Then we additionally explore the phonon-frequency dependent of group velocity $v_\omega$ and relaxation time $\tau_\omega$ in the low frequency region, which plays critical role in determining $\kappa$ since $\kappa = \sum_\omega C_\omega v_\omega^2 \tau_\omega$ with $C_\omega$ being the heat capacity.

Fig. 5(b) shows the frequency dependence of group velocity ($v_\omega$). It is seen that the stacking effect on $v_\omega$ mainly arises in the region of $\omega$<1.5 THz, where the AB' structure has smaller $v_\omega$ than the other two. This result is consistent with the significant downshift of TA and TO' bands along Γ–A path in AB' structure. Note that smaller $v_\omega$ is expected to give rise to a smaller $\kappa$ in AB' structure. Fig. 5(c) additionally shows the frequency dependent phonon relaxation time ($\tau_\omega$). As one can see, the AB' and AA' structures basically have the smallest and largest $\tau_\omega$, respectively. Considering that $\kappa = \sum_\omega C_\omega v_\omega^2 \tau_\omega$, the smaller $v_\omega$ and $\tau_\omega$ in the low-frequency region gives rise to the smallest cross-plane $\kappa$ in AB' structure. On the other hand, although the group velocities of AB and AA' structures are nearly the same, the smaller relaxation time in AB structure makes it have a smaller cross-plane $\kappa$ than AA' structure. This result is consistent with the frequency dependent $\kappa$ of all three structures shown in Fig. 5(a). Note that the stacking-induced variation of $\tau_\omega$ is much more sizeable than that of $v_\omega$, and thus the relaxation time is a dominating factor to the variation of cross-plane $\kappa$.

The stacking effect on phonon relaxation time can be further understood from the variation of TA and TO' branches along Γ–A path. As shown in Figs. 1(d)-1(f), in comparison with the AA' structure, the downshift of TA and TO' branches in AB' structure has two competing effects. On one hand, the downshift of TA band increases the band gap between LA and TA modes and leads to less acoustic-acoustic phonon scatterings, which has an effect of increasing the relaxation time[15]. On the other hand, the downshift of TO' band strengthens the acoustic-optical phonon scatterings, which leads to the decrease of relaxation time. As a result of competition, the additional acoustic-optical scattering effect is more dominating, and the relaxation time in AB' structure is obviously smaller than that of AA' stacking in the low frequency region. This result is confirmed by the relative contributions of different phonon modes on thermal conductivity shown in Fig. 4(d), where the contribution of optical modes in AB' structure is about 5% larger than that in AA' structure. As for the AB stacking structure, both TA and TO' bands slightly upshift in comparison with that of AA' stacking. In this case, although the upshift of TO' band has an effect of reducing the acoustic-optical phonon scattering (the contribution of optical modes becomes 3.4% smaller than that of AA' structure), the increase of acoustic-acoustic phonon scattering induced by the upshift of TA band has a dominating effect on $\kappa$. As a result, the cross-plane $\kappa$ in AB structure is also smaller than that of AA' structure.

Note that the above analysis is based on the bulk h-BN structures, which can explain the stacking effect on cross-plane $\kappa$ with thickness over 1000 nm as shown in Fig. 3(b). Whereas, in the h-BN films with small thickness, the finite size effect would be prominent, resulting in a different stacking dependent cross-plane $\kappa$, as shown in Fig. 3(a) and inset of Fig. 3(b). This is because in the h-BN films with the small thickness (e.g., d<10 nm), much more phonons have MPFs larger than the thickness, which makes them ballistically transport across the h-BN films without scattering. In this case, the ultra-large phonon relaxation time (or MPF) plays a minor role on $\kappa$. Instead, the stacking dependent $\kappa$ would be mainly determined by the variation of the number of excited phonons (denoted as NEP, which is proportional to the heat capacity) for the thermal transport and their group velocities. Since the stacking-induced group velocity difference is small, the variation of $\kappa$ is mainly attributed to the stacking-dependent NEP. This argument is verified by the temperature dependence of cross-plane $\kappa$ with the thickness of d=2.6 nm. As shown in Fig. S1(a), in all the h-BN structures $\kappa$ monotonically increases with temperature increasing from 100 K to 800 K. This result is contrary to that with large thickness, where $\kappa$ decreases with temperature T, following a $\kappa \propto T^{-1}$ law as indicated in Fig. S1(b). It is known that increasing temperature has two effects: (1) it increases the NEP for thermal transport, and (2) it reduces the phonon relaxation time. Fig. S1 means that in the small (large) thickness region, the increase (reduction) of NEP (phonon relaxation time) has dominating effect on cross-plane $\kappa$ of h-BN. Then we additionally calculate the phonon frequency dependent NEP at room temperature, which is obtained by DOS($\omega$)×BE($\omega$), where DOS($\omega$) and BE($\omega$) are the frequency dependent phonon density of states and Bose–Einstein distribution, respectively. As shown in Fig. 5(d), the NEP of AB' structure is significantly larger than that of AA' and AB structures. This result explains why AB' structure has the largest cross-plane $\kappa$ among the three structures in the small thickness region, which is contrary to that in the large thickness region.

**Summary**

In summary, we have explored the stacking and thickness effects on the cross-plane thermal conductivity $\kappa$ of h-BN based on the NEMD simulations and the BTE calculations. We find that $\kappa$ can be significantly modulated by both the stacking structure and thickness d of h-BN. We find that for h-BN films with small thickness region, the variation of κ presents significant ballistic phonon transport characteristics in the cross-plane direction as well as the remarkable stacking effect (κ in AB' structure is up to 60% larger than that in AA'/AB structure). We also find that the stacking effect on κ changes dramatically with thickness increasing, where a crossover appears with d > 200 nm. In the bulk h-BNs, the $\kappa$ of AA' structure becomes about 24%-30% larger than of AB and AB' structures, respectively. We finally clarify that such exotic stacking and thickness dependence of κ is owing to the competing effect of the excited number of phonons and phonon relaxation lifetime, both of which directly affect the thermal conductivity.

S. G. Wang and H. F. Feng contribute equally in this work. The work was supported by the Natural Science Foundation of China (Grant No. 12074301) and Ministry of Science and Technology of the People´s Republic of China (Grant No.

2022YFA1402901). We gratefully acknowledge the computational resources provided by the HPCC platform of Xi'an Jiaotong University.

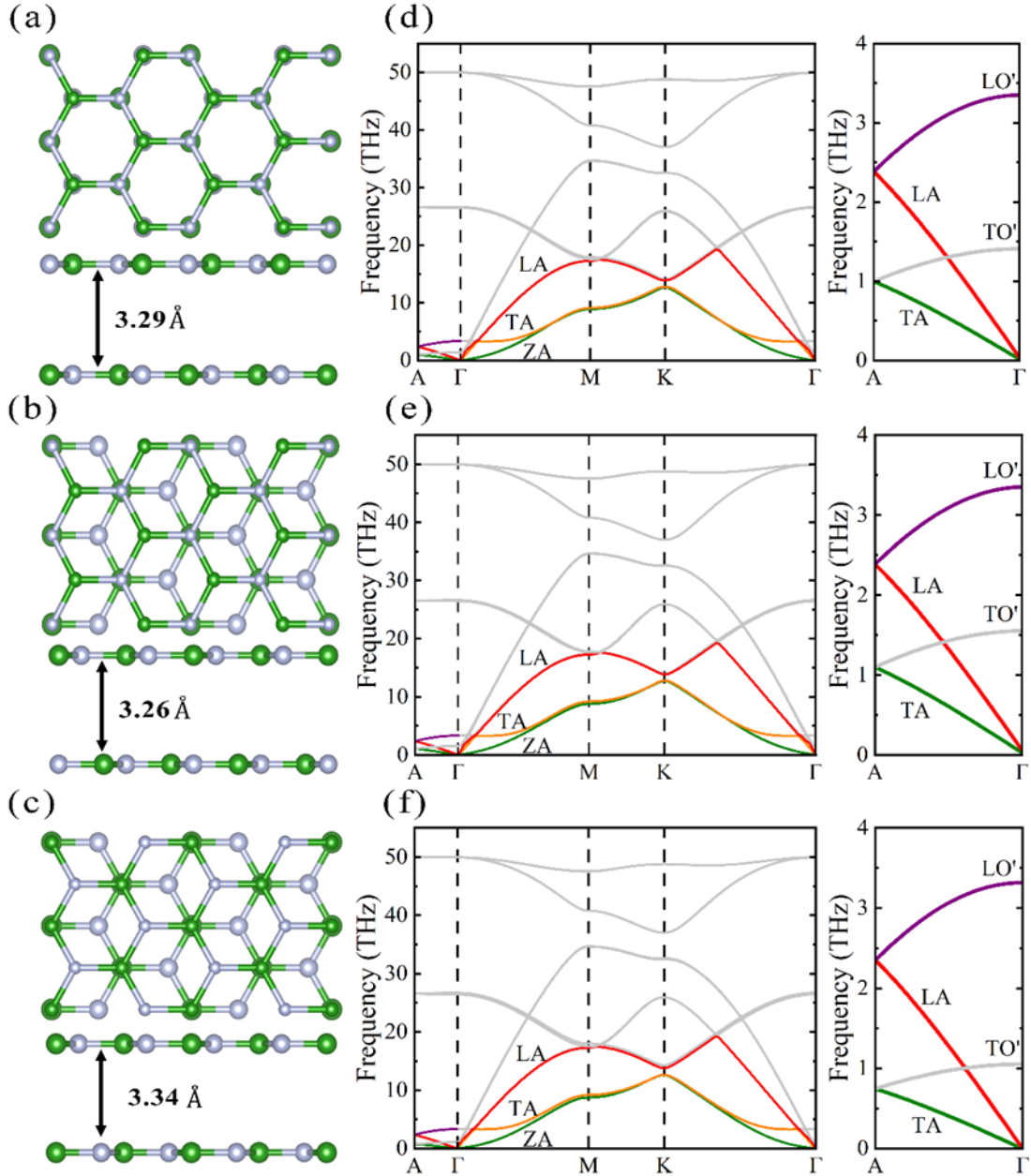

Fig 1. (a)–(c) Top view and side view of h-BN with AA', AB, and AB' stacking structure, respectively. (d)–(f) The corresponding phonon dispersions of h-BN structures in (a)–(c). (a) and (d) The results for AA' stacking h-BN, (b) and (e) represent the results for AB stacking h-BN, and (c) and (f) represent the results for AB' stacking h-BN. In (a)-(c), the green and white balls depict B and N, respectively. The optimized interlayer distance is also shown. In order to facilitate the differentiation of atoms in upper or lower layers in the top view, the radius of the upper atoms is reduced. In (d)–(f), $\Gamma - A$ corresponds to the cross-plane path, and $\Gamma - M - K - \Gamma$ corresponds to the in-plane paths in the Brillouin zone. The right panels in (d)–(f) show the enlarged figure of phonon band structure in $\Gamma - A$ path.

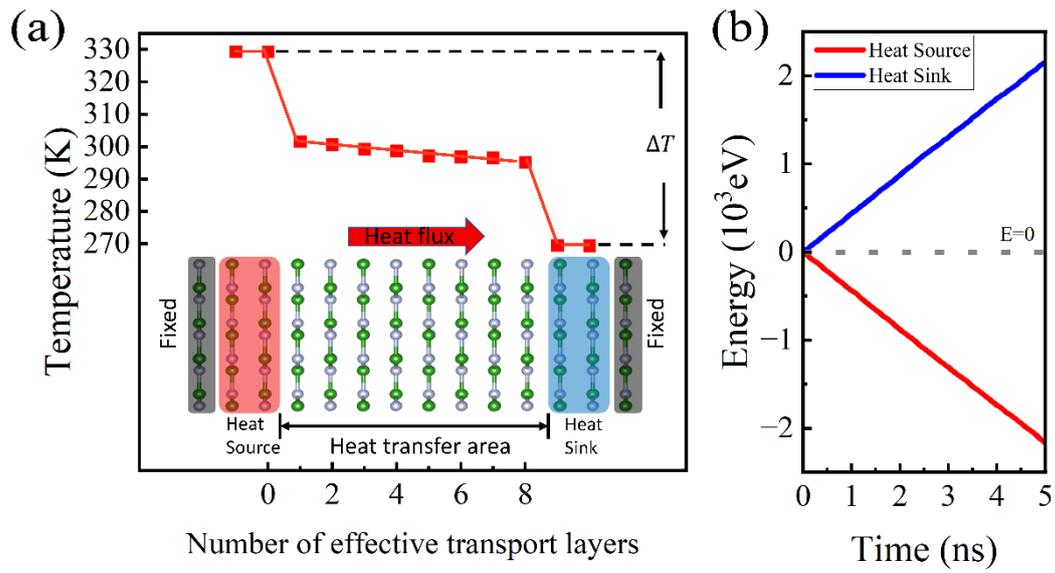

Fig 2. (a) Schematic temperature profile of a h-BN with eight effective transport layers at T=300 K in the NEMD simulation. (b) The corresponding accumulative energy of heat source and heat sink under Langevin thermostat as a function of time.

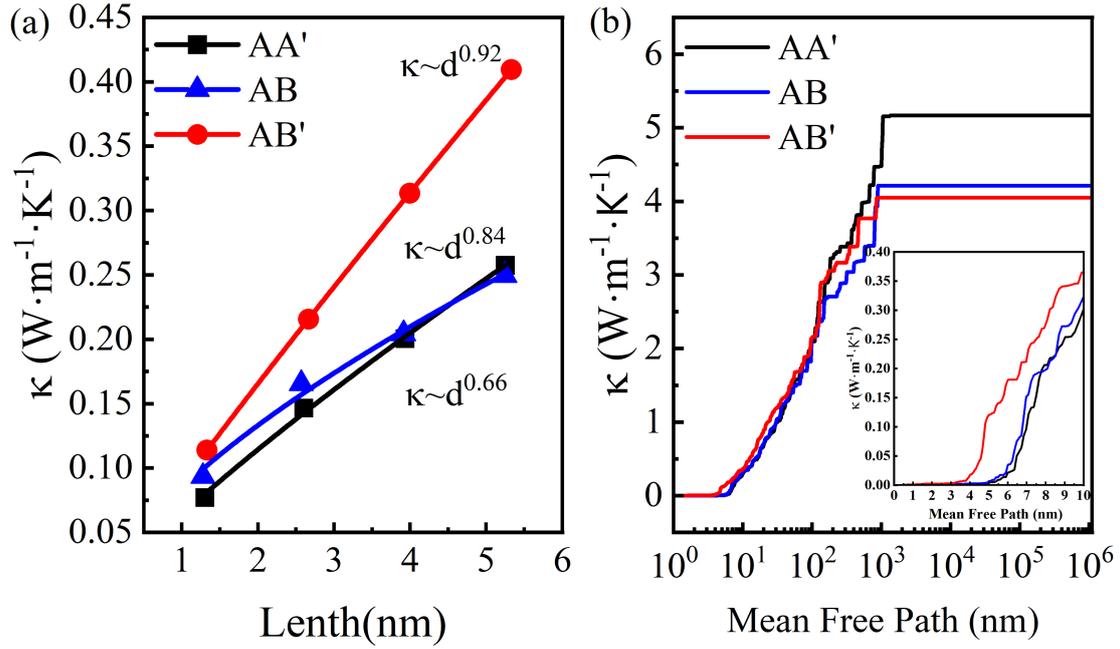

Fig 3. (a) Cross-plane thermal conductivity as a function of thickness at 300 K for all the three h-BN structures. (b) Mean free path dependence of cumulative cross-plane thermal conductivity with AA', AB, AB' stackings at 300 K. The inset shows the enlarged figure for thickness less than 10 nm.

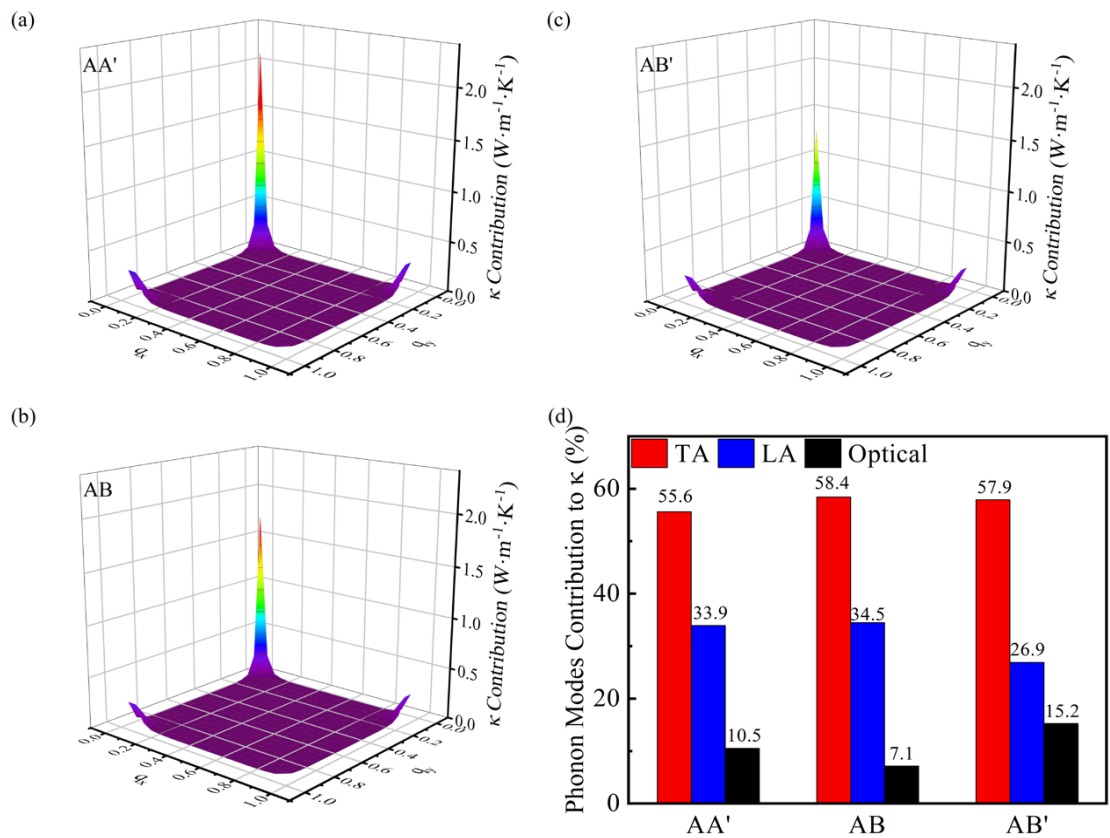

Fig 4. (a)-(c) Cross-plane thermal conductivity projected on the basal plane (0 0 1) in the Brillouin zone for h-BN with AA', AB, and AB' stackings, respectively. (d) Percentage contribution to cross-plane thermal conductivity of phonon modes along the $\Gamma - A$ path for h-BN with AA', AB, and AB' stacking, respectively. The results are obtained at 300 K.

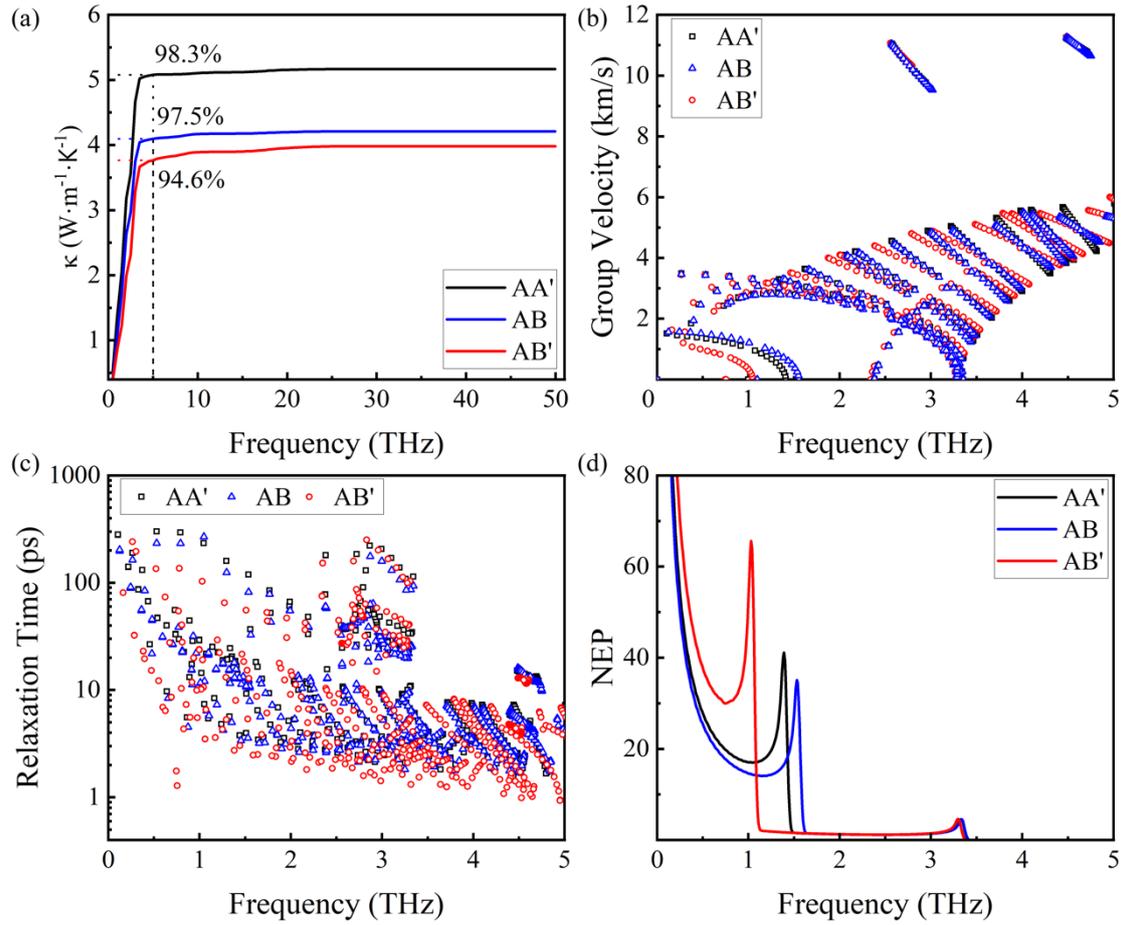

Fig 5. (a) Frequency dependence of cumulative cross-plane thermal conductivity with AA', AB, AB' stackings at 300 K. (b) Group velocity and (c) Phonon relaxation time along the $\Gamma - A$ path at 300 K. The squares, circles, and triangles correspond to the AA', AB, and AB' stacking structure, respectively. (d) Number of excited phonons (NEP) along the $\Gamma - A$ path.